\documentclass{emulateapj}

\usepackage{amsmath,natbib,graphicx,euscript} 

\slugcomment{To appear in ApJ Letters}

\begin{document}

\title{Effective Inner Radius of Tilted Black Hole Accretion Disks}

\author{P. Chris Fragile}
\affil{Department of Physics and Astronomy, College of Charleston,
Charleston, SC 29424}
\email{fragilep@cofc.edu}
\date{{\small    \today}}
\date{{\small   \LaTeX-ed \today}}

\begin{abstract}
One of the primary means of determining the spin $a$ of an astrophysical black hole is by actually measuring the inner radius $r_\mathrm{in}$ of a surrounding accretion disk and using that to infer $a$. By comparing a number of different estimates of $r_\mathrm{in}$ from simulations of tilted accretion disks with differing black-hole spins, we show that such a procedure can give quite wrong answers. Over the range $0 \le a/M \le 0.9$, we find that, for moderately thick disks ($H/r \sim 0.2$) with modest tilt ($15^\circ$), $r_\mathrm{in}$ is nearly independent of spin. This result is likely dependent on tilt, such that for larger tilts, it may even be that $r_\mathrm{in}$ would increase with increasing spin. In the opposite limit, we confirm through numerical simulations of untilted disks that, in the limit of zero tilt, $r_\mathrm{in}$ recovers approximately the expected dependence on $a$.
\end{abstract}

\keywords{accretion, accretion disks --- black hole physics ---
galaxies: active --- X-rays: stars}

\section{Introduction}
\label{sec:intro}
Classical, astrophysical black holes have only two defining properties: mass and angular momentum (or spin). As in other astrophysical contexts, the mass of a black hole can straightforwardly be determined by the application of Kepler's Third Law, provided the black hole is orbited by another observable object and the distance to the system is reasonably well known (at least to within a factor of $\sin i$, where $i$ is the inclination of the binary orbit relative to the observer's line-of-sight). 

The spin $a=Jc/GM$, on the other hand, is a very different matter. So far, no direct measure of spin has been proposed, leaving only indirect means of inferring it. A commonly used method is based on making some measure of the ``inner radius" of the accretion disk. In general relativity, stable circular orbits are not permitted all the way down to the ``surface'' of the black hole, but are instead restricted to lie outside the marginally stable limit $r_\mathrm{ms}$. Beginning with \citet{nov73}, it has commonly been assumed that the accretion disk will observe a similar restriction and truncate at $r_\mathrm{ms}$. Because $r_\mathrm{ms}$ has a well-known monotonic dependence on the spin of the black hole \citep{bar72}, its association with an observed inner radius of the accretion disk is then used to infer the spin of the black hole. The observation of the inner radius can come either from modeling of the continuum emission in the thermally dominant state \citep[e.g.][]{sha06,dav06} or from measuring relativistically broadened reflection features \citep[e.g.][]{wil01}.

However, it has been shown that there are many difficulties with such procedures. First, one has many choices for how to define the effective inner radius $r_\mathrm{in}$ of the accretion disk, and some of these can vary considerably (factors of 2-3) from $r_\mathrm{ms}$ \citep{kro02}. Even if one restricts oneself to such observationally relevant measures as the ``radiation edge'' (the innermost radius from which significant luminosity emerges) or ``reflection edge'' (the smallest radius capable of producing significant X-ray reflection features), significant deviations may be observed \citep{kro02, bec08}. For instance, \citet{ago00} showed that significant emission can emerge from inside $r_\mathrm{ms}$, thus precluding a direct correlation with $r_\mathrm{in}$. Similarly \citet{rey97} showed that the reflection edge is not necessarily tied to $r_\mathrm{ms}$.

Nevertheless, a strong result for untilted disks (disks that lie approximately in the symmetry plane of their black hole spacetimes and have their angular momenta aligned with the spins of their black holes) is that $r_\mathrm{in}$ should gradually decrease with increasing black hole spin, i.e. as $a/M\rightarrow 1$. Therefore, it might at least be hoped that a study of $r_\mathrm{in}$ for a large ensemble of accreting black holes might yield some information about the range and distribution of black hole spins represented in the ensemble.

However, we show in this Letter that there is a further complicating factor. Based on numerical simulations, we show that the effective inner radius of tilted disks (disks that do not have their angular momenta aligned with the spins of their black holes) yield very different results for $r_\mathrm{in}$ than their untilted counterparts for all dynamical measures we have considered. A disk could be tilted for many reasons. In stellar mass binaries, the
orientation of the outer disk is fixed by the binary orbit, whereas the orientation of the black hole is determined by how it became part of the system. If the black hole formed from a member of a preexisting binary through a supernova, then the black hole could be tilted if the explosion were asymmetric. If the black hole joined the binary through multi-body interactions, such as binary capture or replacement, then there would have been no preexisting symmetry, so the resulting system would nearly always harbor a tilted black hole. This same argument can be extended to AGN in which merger events reorient the central black hole or its fuel supply and result in repeated tilted configurations. Tilted disks are subject to additional torques due to differential Lense-Thirring precession \citep{bar75}.

Our results indicate that for moderately thick ($H/r \sim 0.2$), tilted ($15^\circ$) disks, $r_\mathrm{in}$ is independent of spin, or, in some measures, actually {\em increases} with increasing black hole spin. These results could have important consequences for observational efforts to constrain black hole spins.

\section{Numerical Simulations}
\label{sec:numerics}
The numerical data in this Letter were taken from simulations presented in \citet{fra07} and \citet{fra09}, as well as two previously unpublished simulations. All of the simulations used the Cosmos++ GRMHD code \citep{ann05}. The simulations presented in \citet{fra07} and the two previously unpublished simulations used a spherical-polar grid in Kerr-Schild coordinates, with an effective grid resolution of $128^3$, except near the poles which were purposefully underresolved. The simulations from \citet{fra09} used either this same spherical-polar grid or a ``cubed-sphere'' grid also in Kerr-Schild coordinates, with a resolution of $128\times64\times64$ on each of 6 logically connected blocks that are morphed into segments of a sphere. As such, all of these simulations used what was referred to as ``high'' spatial resolution in \citet{fra07} and \citet{fra09}. In all cases, the simulations were initialized with an analytically solvable, time-steady, axisymmetric gas torus \citep{dev03}, threaded with a weak poloidal magnetic field with minimum $P_{gas} /P_{mag} = 10$ initially. The magnetorotational instability (MRI) arose naturally from the initial conditions. The simulations were all evolved for $\sim7900M$, or $\sim40$ orbits at $r = 10M$ in units with $G = c = 1$. Only data from the final $\sim790M$ of the simulation are used in this Letter, ensuring that the disk is fully turbulent and transient effects from the initial conditions have died down. The principle variables that are studied in these simulations are the spin of the black hole, varying over the range $0 \le a/M \le 0.9$, and the initial tilt between the disk and the black hole, being either $0$ or $15^\circ$. The different simulations are summarized in Table \ref{tab:params}.

\begin{deluxetable}{cccc}
\tabletypesize{\scriptsize}
\tablecaption{Simulation Parameters \label{tab:params}}
\tablewidth{0pt}
\tablehead{
\colhead{Simulation} & \colhead{$a/M$} & \colhead{Tilt} &
\colhead{Grid} \\
\colhead{} & \colhead{} & \colhead{Angle} & 
\colhead{} 
}
\startdata
0H\tablenotemark{a} & 0 & ... & Spherical-polar \\
315H & 0.3 & $15^\circ$ & Spherical-polar \\
50H\tablenotemark{a} & 0.5 & $0^\circ$ & Cubed-sphere \\
515H-S\tablenotemark{a} & 0.5 & $15^\circ$ & Spherical-polar \\
515H-C\tablenotemark{a} & 0.5 & $15^\circ$ & Cubed-sphere \\
715H & 0.7 & $15^\circ$ & Spherical-polar \\
90H\tablenotemark{b} & 0.9 & $0^\circ$ & Spherical-polar \\
915H\tablenotemark{b} & 0.9 & $15^\circ$ & Spherical-polar
\enddata
\tablenotetext{a}{\citet{fra09}}
\tablenotetext{b}{\citet{fra07}}

\end{deluxetable}

\section{Effective Inner Radius}
\label{sec:rin}
As pointed out in \citet{kro02}, there are many possible definitions of $r_\mathrm{in}$. In this Letter we will ignore any that rely on knowledge of the radiation field, instead sticking with definitions based on dynamical fluid properties that are easily obtained directly from the simulations. We explore four such definitions.

To best represent the quasi-steady states achieved in the simulations, all data are time-averaged over the interval $t_{min} \approx 7100M$ to $t_{max} \approx 7900M$. To ensure all derived quantities are representative of the disk and not the low-density background gas, we apply density-weighted shell averaging, i.e. we present quantities of the form 
\begin{align}
\langle\mathcal{Q}\rangle_i = \frac{1}{\langle \rho \rangle_i} \int^{t_{max}}_{t_{min}} \int^{2\pi}_0
\int^\pi_0 &\rho (r_i, \theta, \phi, t) \times \\ \nonumber
 &\mathcal{Q}(r_i,\theta, \phi, t) \sqrt{-g}
\, \mathrm{d}\theta \mathrm{d}\phi \mathrm{d}t ~,
\end{align}
where 
\begin{equation}
\langle \rho \rangle_i = \int^{t_{max}}_{t_{min}} \left[ \frac{\int^{2\pi}_0 \int^\pi_0 \rho (r_i, \theta, \phi, t) \sqrt{-g} \mathrm{d}\theta \mathrm{d}\phi}{\int^{2\pi}_0 \int^\pi_0 \sqrt{-g} \mathrm{d}\theta \mathrm{d}\phi} \right] \mathrm{d}t
\end{equation}
is the average density of a given radial shell $i$. 

\subsection{Surface Density}
\label{sec:density}
Perhaps the conceptually simplest definition of $r_\mathrm{in}$ is where the surface density of the disk drops significantly from some relevant peak value, especially if this occurs over some relatively narrow radial range. Indeed, we generically find that the surface density drops rapidly from a local peak at around $r \sim 10M$
[see e.g. Figure 13 of \citet{fra07}]. For concreteness we take $\Sigma(r_\mathrm{in}) = \Sigma_\mathrm{max} /3e$, where $\Sigma(r_i) \equiv \langle \rho \rangle_i \langle H \rangle_i$ and $\Sigma_\mathrm{max} = \mathrm{max}[\Sigma(r_i)]$ within the interval $r_{BH} \le r_i \le 20M$. The local scale height of the disk is estimated from 
\begin{equation}
H = \frac{\rho}{\vert \nabla \rho \vert} ~.
\label{eqn:H}
\end{equation}
Note that this definition differs from what we used in \citet{fra07} and gives slightly larger values for $H$, though the resulting profile is quite similar. 

\begin{figure}
\plotone{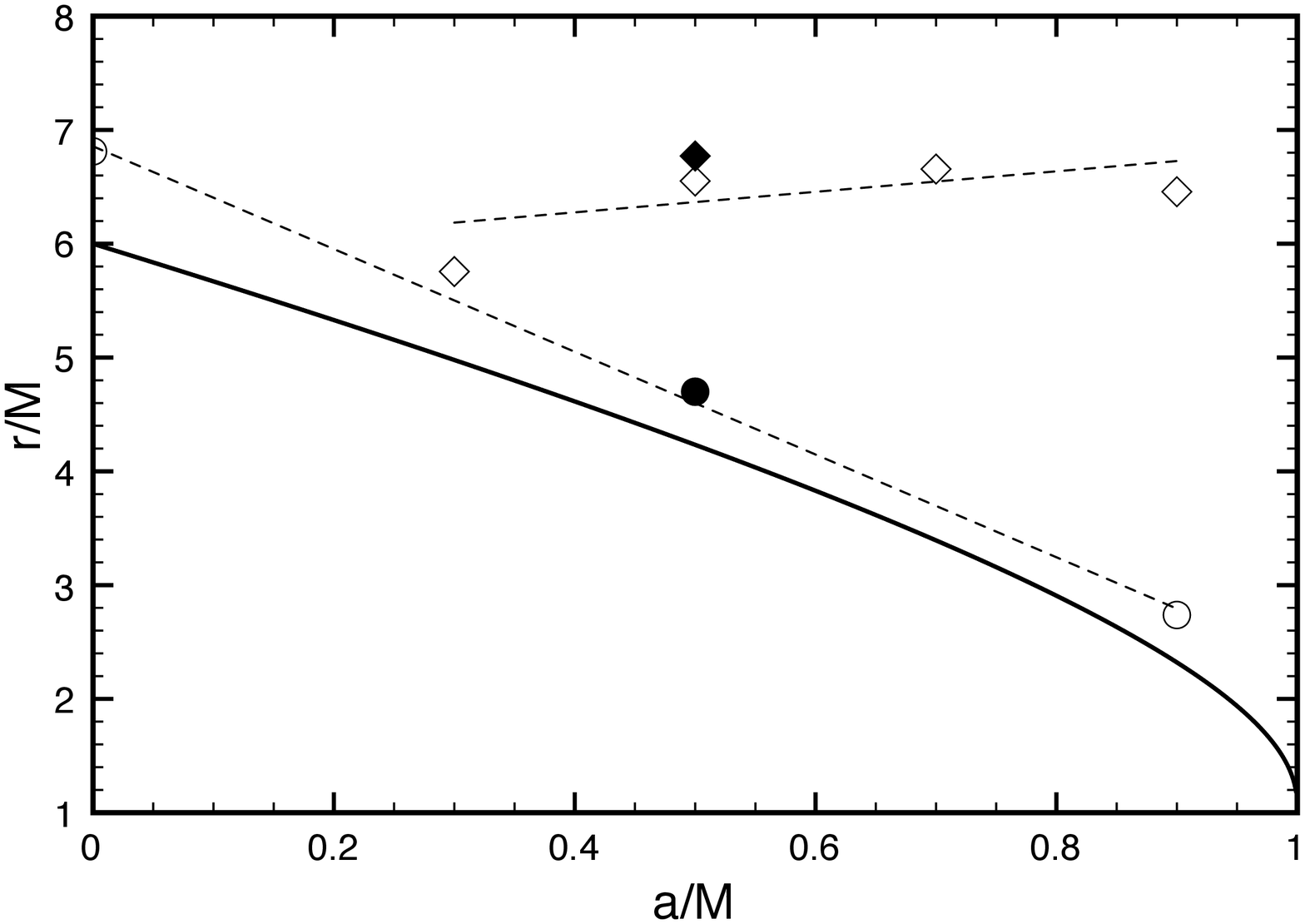}
\caption{Plot of the effective inner radius $r_\mathrm{in}$ of simulated untilted ({\em circles}) and tilted ({\em diamonds}) accretion disks as a function of black-hole spin using the surface density measure $\Sigma(r_\mathrm{in}) = \Sigma_\mathrm{max}/3e$. {\em Open} symbols use the spherical-polar grid from \citet{fra07} and {\em filled} symbols use the cubed-sphere grid from \citet{fra09}. The {\em solid} line is the marginally stable orbit $r_\mathrm{ms}$, and the light, dashed lines are illustrative trend lines of the untilted and tilted data.
\label{fig:density}}
\end{figure}

In Figure \ref{fig:density} we plot the resulting estimates of $r_\mathrm{in}$ from each simulation. For illustration purposes, we also plot a line representing the value of $r_\mathrm{ms}$ as a function of spin. The most obvious result of this figure is that, although the untilted simulations ({\em circles}) follow the trend indicated by $r_\mathrm{ms}$ (though with systematically higher numerical values), the tilted simulations ({\em diamonds}) do not. In fact, instead of $r_\mathrm{in}$ decreasing with increasing black-hole spin, it {\em increases} slightly. We further emphasize this difference by fitting simple trend lines to the untilted and tilted data separately. Clearly the data are following two very different trends. This is crucial since it means that any estimates of $a/M$ based on the tilted data could be very wrong. For instance, the 915H simulation (with a spin of $a/M=0.9$ and a tilt of $15^\circ$) produces a value for $r_\mathrm{in}$ that is almost identical to the 0H ($a/M=0$) simulation.

\subsection{Inflow Time}
\begin{figure}
\plotone{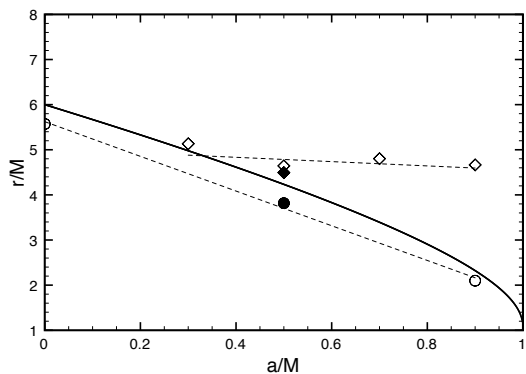}
\caption{Plot of the effective inner radius $r_\mathrm{in}$ of black-hole spin using the inflow time measure $t_\mathrm{in}(r_\mathrm{in}) = 3t_\mathrm{orb}(r_\mathrm{in})$. The symbols have the same meanings as in Fig. \ref{fig:density}.
\label{fig:inflow}}
\end{figure}

Another possible definition for $r_\mathrm{in}$ is where the radial infall time $t_\mathrm{in}(r_i) = r_i/\langle {V}^r \rangle_i$ of material moving in through the disk becomes shorter than some factor of the local orbital time $t_\mathrm{orb}(r) = 2\pi/\Omega$, where $\Omega = (M/r^3)^{1/2}/[1+a(M/r^3)^{1/2}]\approx\Omega_{\rm Kep}$ is the local orbital angular frequency.
Specifically, if we take $t_\mathrm{in}(r_\mathrm{in}) = 3t_\mathrm{orb}(r_\mathrm{in})$, then Figure \ref{fig:inflow} shows the resulting estimates of $r_\mathrm{in}$. The factor of 3 in our definition of $t_\mathrm{in}(r_\mathrm{in})$ is rather arbitrary [\citet{kro02} used a value of 10], and is simply chosen for convenience. Clearly if a smaller coefficient is used, all of the data points in Figure \ref{fig:inflow} would shift to smaller radii, closer to the event horizon; whereas if a larger coefficient is used, all of the data points would move to larger radii. A similar caveat applies to all of the normalizations used in this Letter. The key point is that the trend lines would remain largely unchanged, and it is the trend lines we are most interested in, not specific numerical values of $r_\mathrm{in}$.

Looking at the trend lines in Figure \ref{fig:inflow}, we again find with this measure of $r_\mathrm{in}$ that the untilted simulations closely follow the prediction of $r_\mathrm{ms}$. The tilted simulations again deviate from this trend; instead they now show an almost flat trend line. By this measure, $r_\mathrm{in}$ for tilted disks appears to be {\em independent of spin}. Of course, this again means that simulation 915H produces a value for $r_\mathrm{in}$ that is very close to the 0H simulation. 

\subsection{Turbulence Edge}
\begin{figure}
\plotone{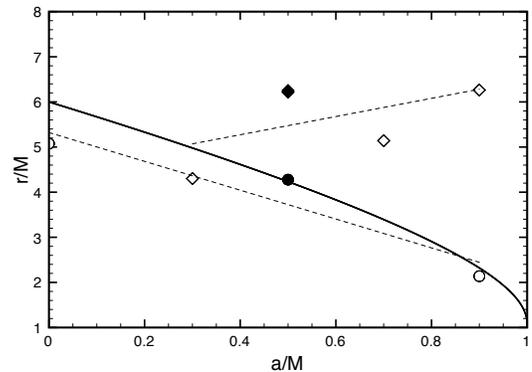}
\caption{Plot of the effective inner radius $r_\mathrm{in}$ of black-hole spin using the turbulence edge measure $(V_\mathrm{turb}/\langle{V}^r\rangle)(r_\mathrm{in}) = (V_\mathrm{turb}/\langle{V}^r\rangle)_\mathrm{max}/2e$. The symbols have the same meanings as in Fig. \ref{fig:density}. Note that the values for simulations 515H-S and 515H-C nearly overlap.
\label{fig:turb}}
\end{figure}

\citet{kro02} suggested another possible definition of $r_\mathrm{in}$ as the radius where the local turbulent velocity $V_\mathrm{turb}$ becomes short compared to the local infall velocity $\langle {V}^r \rangle$.  Normally in the bulk of the disk, MRI driven turbulence produces velocity fluctuations that are substantially larger on average than the mean infall velocity. However, as disk material approaches the black hole, its radial infall velocity increases more rapidly than its turbulent velocity. For this measure we define the inner radius as $(V_\mathrm{turb}/\langle{V}^r\rangle)(r_\mathrm{in}) = (V_\mathrm{turb}/\langle{V}^r\rangle)_\mathrm{max}/2e$, where $(V_\mathrm{turb}/\langle{V}^r\rangle)_\mathrm{max}$ is defined over the interval $r_{BH} \le r_i \le 15M$. 
For our purposes we take $V_\mathrm{turb}(r_i) \equiv {V}^r_\mathrm{rms}(r_i)$, where
\begin{equation}
V^r_\mathrm{rms}(r_i) = \sqrt{ \langle (V^r)^2 \rangle_i } ~.
\end{equation}
We see in Figure \ref{fig:turb} that this definition of $r_\mathrm{in}$ gives similar results to the previous two. Again, $r_\mathrm{in}$ follows the expected trend for the untilted simulations, but appears to increase with spin for the tilted ones.

\subsection{Magnetic Field Structure}
\begin{figure}
\plotone{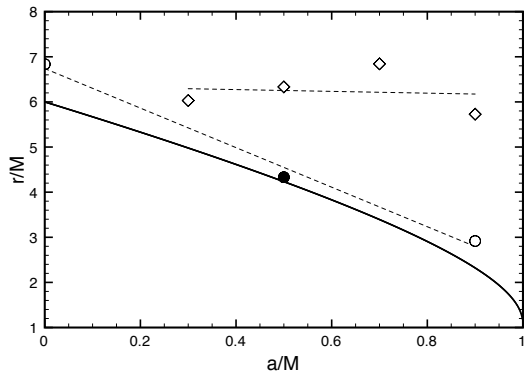}
\caption{Plot of the effective inner radius $r_\mathrm{in}$ of black-hole spin using the magnetic field structure measure $\alpha_\mathrm{mag}(r_\mathrm{in}) = 2e (\alpha_\mathrm{mag})_\mathrm{min}$. The symbols have the same meanings as in Fig. \ref{fig:density}.
\label{fig:alpha}}
\end{figure}

As a final possible definition of $r_\mathrm{in}$, we consider the radius where the local magnetic field structure transitions from MRI-driven turbulence toward an evolution better described as flux-freezing. Although a purely turbulent process would not lead to any correlation, the linear properties of the MRI and the consistent orbital shear of the disk impose some modest correlations between $B^r$ and $B^\varphi$, such that $\alpha_\mathrm{mag} \sim 0.001$ \citep{kro02}, where
\begin{equation}
\alpha_\mathrm{mag} = \frac{ \vert u^r u^\varphi \vert B \vert^2 - B^r B^\varphi \vert}{4 \pi P}
\end{equation}
is the dimensionless magnetic stress parameter. As the flux-freezing limit is approached, correlations between the magnetic field components become stronger, and $\alpha_\mathrm{mag} \rightarrow 1$. In this region, the transfer of energy to smaller scales becomes slower than the infall time. For our purposes, we take $\alpha_\mathrm{mag} (r_\mathrm{in}) = 2e (\alpha_\mathrm{mag})_\mathrm{min}$,  where $(\alpha_\mathrm{mag})_\mathrm{min}$ is defined over the interval $r_{BH} \le r_i \le 20M$. 
We see in Figure \ref{fig:alpha} that this definition of $r_\mathrm{in}$ gives similar results to the previous ones. Again, $r_\mathrm{in}$ follows the expected trend for the untilted simulations, but appears to be nearly independent of spin for the tilted ones.

Note in Figure \ref{fig:alpha} no datum represents the cubed-sphere simulation 515H-C. This is because it is very difficult on the cubed-sphere grid to properly recover the $B^r$ and $B^\phi$ components, especially close to the black hole for tilted disks. Therefore, we were not able to recover a meaningful value for $\alpha_\mathrm{mag}$ much inside $\sim 8M$ for that simulation.

\section{Discussion}
\label{sec:discussion}
We have explored four possible measures of the effective inner radius of simulated black-hole accretion disks, based on surface density, inflow and turbulent velocities, and magnetic field structure. With all four measures, we found two consistent trends: 1) for untilted simulations $r_\mathrm{in}$ closely follows the analytic form of $r_\mathrm{ms}$; whereas 2) for $15^\circ$ tilted simulations $r_\mathrm{in}$ shows a flat or even increasing trend as $a/M\rightarrow 1$. 

The fact that Figures \ref{fig:density} - \ref{fig:alpha} all show remarkable similarities suggests that our results are not due to a poor choice of definition for $r_\mathrm{in}$. Also, although we have ignored the dependence of $r_\mathrm{ms}$ on inclination in making our plots, Figure 5 of \citet{fra07} shows that the deviation would be quite small for $i = 15^\circ$, as would be appropriate for this work.

We can get some estimate of the uncertainties associated with our results by comparing data obtained from the two 515H simulations, one done on the spherical-polar grid and one on the cubed-sphere. For the first three measures of $r_\mathrm{in}$, where we were able to get estimates using both simulations, they agree remarkably well. Furthermore, for most of the measures the trend lines fit the data well, suggesting small uncertainties. 

We conclude that the discrepancy of $r_\mathrm{in}$ between tilted and untilted simulations is a robust result, and must, therefore, be rooted in some physical process. The most likely agent of such a process would seem to be the standing shocks aligned along the line of nodes between the black hole symmetry plane and disk midplane \citep{fra08}. These features are not present in simulations of untilted disks. Such non-axisymmetric shocks can be quite efficient at extracting angular momentum from the gas flow, which would reduce the effect of the black hole spin on $r_\mathrm{in}$. 

In this work we have only considered two values of initial tilt: $\beta_i = 0$ and $15^\circ$. It stands to reason, however, that as $\beta_i \rightarrow 0$ the inner radius would approach the untilted values obtained in this work. On the other hand, for $\beta_i > 15^\circ$, it may well be that the discrepancies in $r_\mathrm{in}$ would be even larger. In such a case, $r_\mathrm{in}$ might truly increase with increasing spin.

The results in this Letter then imply that, at least for moderately thick accretion disks ($H/r \gtrsim 0.1$), measurements of $r_\mathrm{in}$ are {\em not} reliable predictors of $a$, unless one can independently confirm that the disk is not tilted. This could pose a problem for recent attempts to determine spin using so-called ``Hard'' state observations \citep{mil06, rei08}, where the flow is generally taken to be thick. 

For very thin disks ($H/r \lesssim 0.01$), the effect of tilt is expected to be much different. Such disks are expected to be subject to the Bardeen-Petterson effect \citep{bar75}, where the midplane of the disk at small radii would be aligned with the symmetry plane of the black hole through the competing action of Lense-Thirring precession and viscous diffusion, while possibly leaving an outer disk that is tilted at large radii. Since the inner disk would be aligned with the symmetry plane of the black hole, the effective inner radius should be the same as for an untilted disk. This would apply for disks in the ``Soft'' or ``Thermally-Dominant'' state, with luminosities well below the Eddington limit, which are likely to be thin. Attempts to estimate the black hole spin by modeling the disk in this state may be unaffected by the cautions raised in this Letter.  However, for sources with disk luminosities near or above the Eddington limit \citep[e.g.][]{mid06}, the disk should again be thick, and the effect discussed here will apply.  Even if the disk is thin, we reiterate that it is still important to independently fit for the inclination of the inner disk, since this may not be the same as the binary inclination. 

Other methods of determining black-hole spin, such as using quasi-periodic oscillation (QPO) frequencies, may not be affected by our results. For instance, QPO models based on resonant frequencies attributable to the black hole spacetime itself \citep[e.g.][]{abr01} would likely not be strongly affected. In fact, the weak dependence of $r_\mathrm{in}$ on $a$ detailed in this Letter actually helps one model of low-frequency QPOs based on the precession of a radially extended thick disk, by ensuring that the maximum frequency would not exceed $\sim 10$ Hz regardless of black hole spin, as observed \citep{ing09}. However, to date, no consensus model for QPOs has emerged, leaving us with few options for unambiguously determining black-hole spin.

To end, we emphasize that it would be wrong to conclude from this paper that all tilted black hole accretion disks should appear to have $r_\mathrm{in} \approx 6 M$. Remember, the normalizations used in each of our plots were chosen simply to give values for the untilted disks that were reasonably similar to the values for $r_\mathrm{ms}$. There is no other physical motivation for choosing those normalizations, and a different choice would have led to different numerical values for $r_\mathrm{in}$. Further, we can not really say how the dynamically determined $r_\mathrm{in}$ in this work would compare to values of $r_\mathrm{in}$ measured from continuum modeling or iron-line fitting without doing more work. Specifically, we would need to generate synthetic disk images from the simulation data, and measure a radiation or reflection edge. Such work is already underway using the relativistic ray tracing code of \citet{dex09}. Even then, there is the problem that the original simulations did not properly account for all of the important physical processes occurring in the disk, notably heating due to dissipation of turbulent energy and radiative cooling. That will have to await future simulations. 

What {\em can} be concluded from this paper is that all tilted disks (with a tilt around $15^\circ$) should appear to have the {\em same} $r_\mathrm{in}$, independent of spin. Our results may further imply that measurement of a small value for $r_\mathrm{in}$ (without defining small) would require a rapidly rotating black hole {\em and} an untilted disk, whereas a large value of $r_\mathrm{in}$ would {\em not} necessarily mean that the black hole is spinning slowly; it could simply be tilted. 

\begin{acknowledgements}
I would like to thank Sam Cook for his help in completing the new simulations presented in this Letter and the preparation of this manuscript. I thank Jason Dexter, Shane Davis, and the anonymous referee for very useful feedback on this work. I gratefully acknowledge the support of a REAP grant from the South Carolina Space Grant Consortium. This work was supported by the National Science Foundation under grant AST 0807385 and through TeraGrid resources provided by the Texas Advanced Computing Center (TACC). This work also made use of computing resources provided by the Barcelona Supercomputing Center under activity AECT-2007-3-0002.
\end{acknowledgements}


\end{document}